# Non-degenerate Ground State of the Spin-Boson Model under Abelian Diagonalization

Tao Liu[1,*], Juhao Wu[2], Make Ying[3]

[1]*Southwest University of Science and Technology, Mianyang, Sichuan 621010, China*
[2]*Stanford University, Stanford, California 94309, USA*
[3]*EXTEC Inc*

liut@swust.edu.cn, juhaowu@stanford.edu, Lightmakejlc@gmail.com

**Abstract** By utilizing a unitary transformation, we derive the necessary and sufficient conditions for the degeneracy between the even- and odd-parity energy states of the spin-boson model (SBM). Employing the Rayleigh quotient of matrix algebra, we rigorously prove that the ground state energy of the SBM is lower than the systems lowest possible degenerate energy and possesses a definite parity. Based on the necessary and sufficient conditions for parity breaking, we provide an analytical expression for the parity-breaking critical value, which is closely related to the expansion order and computational accuracy. This expression reproduces the SBM phase diagram obtained by quantum Monte Carlo (QMC) and logarithmic discretization numerical renormalization group (NRG) methods. However, this phase diagram does not characterize the ground state of the system.

**I. Introduction**

The coupling between a two-level system and its environment provides a unique platform for verifying fundamental principles of interesting quantum physics. The phenomenological description of an open quantum system, where the environment is modeled by multi-mode harmonic oscillators—the spin-boson model—plays a significant role in quantum information and condensed matter physics. Particularly, the quantum phase transition (QPT) between the spin-localized phase and the delocalized phase at zero temperature in the SBM has attracted considerable interest.

The standard SBM Hamiltonian with zero local field (in units of ℏ = 1) is given by [1],

$$H_{sb} = -\frac{\Delta}{2}\sigma_x + \sum_k \omega_k a_k^\dagger a_k + \frac{1}{2}\sum_k \lambda_k(a_k^\dagger + a_k)\sigma_z. \tag{1}$$

where $\sigma_x$ and $\sigma_z$ are the usual Pauli operators, $\Delta$ is the tunneling amplitude between the two levels of the spin. $a_k^\dagger$ and $a_k$ are the creation and annihilation operators of the bath modes with frequencies ω_k, and λ_k is the coupling strength between the spin and the bath mode k. The effect of the harmonic oscillator environment is reflected by the spectral function $J(\omega) = \pi\sum_k \lambda_k^2 \delta(\omega - \omega_k)$ for $0 < \omega < \omega_c$, with the cutoff energy $\omega_c$. [The cutoff $\omega_c$ originates from the conventional treatment of Ohmic dissipation, implying vanishing high-frequency contributions.] In the infrared limit, i.e., $\omega \to 0$, the power-law behavior of $J(\omega)$ is of particular importance. Considering the low-energy details of the spectrum, we have $J(\omega) = 2\pi\alpha\omega_c^{1-s}\omega^s$ for $0 < \omega < \omega_c$ and dissipation strength α [1]. The exponent s characterizes different bath types: super-Ohmic bath for s > 1, Ohmic bath for s = 1, and sub-Ohmic bath for s < 1.

Typically, the QPT is believed to occur in the absence of the local field (ε=0) [Note: For the SBM with a local field, the system lacks parity symmetry, and the phase transition is usually not discussed in this context.] [2–4]. It is thought to transition from a non-degenerate ground state with zero magnetization below a critical coupling to a twofold-degenerate ground state with finite magnetization for couplings larger than the critical value. This QPT has been reported by various numerical studies, such as the numerical renormalization group (NRG) technique [2-3, 5-6, 7-11], quantum Monte Carlo (QMC) [4], exact diagonalization (ED) [12], the density matrix renormalization group (DMRG) approach [13], and the variational matrix product state (VMPS) approach [14]. Analytical studies, such as the unitary transformation method [15, 16], have also been employed. Recent work using variational methods also reported the appearance of the QPT between delocalization and localization [17] for ε = 0.

However, all evidence so far indicating the emergence of ground state degeneracy in the SBM with increasing dissipation strength comes from the practical computational results of various approximate methods [Note: Even analytical approximations stem from finite expansions, e.g., Ref. [17]. There has never been a rigorous analytical proof that the SBM ground state necessarily becomes degenerate.]. A fundamental question long overlooked is: **Is the emergence of ground state degeneracy an intrinsic property of the system itself, or is it an artifact caused by the approximations inherent in computational methods?** Due to the high degrees of freedom of the open quantum system modeled by multi-mode harmonic oscillators, verifying this problem through finite numerical calculations faces enormous difficulties on the one hand. On the other hand, the ground state problem of the model, in principle, reduces to solving a general algebraic equation of degree much higher than four. According to Abels theorem, such equations generally have no algebraic solution. Thus, this fundamental problem seems unsolvable. Let us re-examine this issue from a different perspective: A promising idea is that **if we could analytically obtain the degenerate energies of the SBM and theoretically determine the relationship between the ground state energy and the lowest possible degenerate energy, we could provide a definitive answer to this problem.** Fortunately, due to the parity symmetry of the SBM Hamiltonian under ε = 0, the complete set of all possible degenerate energies between the even- and odd-parity energy spectra can be obtained analytically.

This paper first starts from Eq. (1) and derives two Hamiltonians separated by parity via a unitary transformation; from this, we derive the necessary and sufficient conditions for degeneracy in the parity energy spectra and the analytical form of the degenerate energies. Furthermore, **using the Rayleigh quotient of matrix algebra,**

**we rigorously prove that the minimum energy of the SBM is lower than the lowest degenerate energy of the even- and odd-parity spectra.**

Secondly, based on the necessary and sufficient conditions for parity breaking, we strictly derive an analytical expression for the parity-breaking critical value, which is closely related to computational accuracy and expansion order. The quantum phase transition critical points obtained in the past by quantum Monte Carlo (QMC) and logarithmic discretization numerical renormalization group (NRG) methods are highly consistent with the results of this analytical expression. **This so-called quantum phase transition of the SBM is not a signature of ground state parity breaking. The ground state of the system possesses a definite parity and is non-degenerate.**

## II. Unitary Transformation of Hamiltonian and Time-Independent Schrödinger Equation

For Hamiltonian (1), the parity operator $P_{sb} = \sigma_x e^{i\pi \sum_k a_k^\dagger a_k}$ commutes with $H_{sb}$ ($[P_{sb}, H_{sb}] = 0$). Introducing the unitary transformation $U = \frac{1}{\sqrt{2}}\left[[1, e^{-i\pi \sum_k a_k^\dagger a_k}], [-1, e^{i\pi \sum_k a_k^\dagger a_k}]\right]$, we obtain $P = U P_{sb} U^\dagger = \left(e^{i\pi \sum_k a_k^\dagger a_k}\right)^2 \sigma_z = \sigma_z$, and:

$$H = U H_{sb} U^{-1} = \begin{pmatrix} H^+ & 0 \\ 0 & H^- \end{pmatrix} \tag{2}$$

$$H^+ = H_0 - H_\Delta \tag{3}$$

$$H^- = H_0 + H_\Delta \tag{4}$$

where $H_0 = \sum_k \omega_k (A_k^\dagger A_k - q_k^2)$, $H_\Delta = \Delta e^{i\pi \sum_k a_k^\dagger a_k}/2$, $A_k^\dagger = a_k^\dagger + q_k$, $A_k = a_k + q_k$, $q_k = \lambda_k/2\omega_k$. The transformation decomposes the spectrum into two independent parity sectors with energies $E^\pm$.

Assuming $|\psi\rangle = [[|\phi^+\rangle], [|\phi^-\rangle]]$, we have $P[[|\phi^+\rangle], [0]] = [[|\phi^+\rangle], [0]]$, $P[[0], [|\phi^-\rangle]] = -[[0], [|\phi^-\rangle]]$, and $H^\pm|\phi^\pm\rangle = E^\pm|\phi^\pm\rangle$. Thus $H^+$ and $H^-$ are independent Hamiltonians for even and odd parity subspaces. The wavefunctions satisfy:

$$|\phi^+\rangle = \sum_{\{n\}} c^+_{\{n\}} |\{n\}\rangle_A \tag{5}$$

$$|\phi^-\rangle = \sum_{\{n\}} c^-_{\{n\}} |\{n\}\rangle_A \tag{6}$$

where $\{n\} = n_1, \cdots, n_N$ (N modes), $|\{n\}\rangle_A = \Pi_{k=1}^N |n_k\rangle_{A_k}$ with

$$|n_k\rangle_{A_k} = \frac{(a_k^\dagger + q_k)^{n_k}}{\sqrt{n_k!}} e^{-q_k a_k^\dagger - \frac{q_k^2}{2}}|0\rangle$$

$$_A\langle\{m\}|\{n\}\rangle_A = \delta_{\{m\},\{n\}}$$

Substituting (3)-(6) into the Schrödinger equation yields:

$$\sum_k \omega_k [A_k^\dagger A_k - q_k^2] |\phi^+\rangle - \frac{\Delta}{2} e^{i\pi \sum_k a_k^\dagger a_k}|\phi^+\rangle = E^+|\phi^+\rangle \qquad (7)$$

$$\sum_k \omega_k [A_k^\dagger A_k - q_k^2] |\phi^-\rangle + \frac{\Delta}{2} e^{i\pi \sum_k a_k^\dagger a_k}|\phi^-\rangle = E^-|\phi^-\rangle \qquad (8)$$

Since $A_k^\dagger A_k |\{n\}\rangle_A = n_k |\{n\}\rangle_A$, we have:

$$\sum_k \omega_k (m_k - q_k^2) |\phi^+\rangle - \frac{\Delta}{2} e^{i\pi \sum_k a_k^\dagger a_k}|\phi^+\rangle = E^+|\phi^+\rangle \qquad (9)$$

$$\sum_k \omega_k (m_k - q_k^2) |\phi^-\rangle + \frac{\Delta}{2} e^{i\pi \sum_k a_k^\dagger a_k}|\phi^-\rangle = E^-|\phi^-\rangle \qquad (10)$$

## III. Boson Parity Breaking and Energy Level Degeneracy

Since $[H_0, H_\Delta] \neq 0$, $H_0$ and $H_\Delta$ share no common eigenstates in $\{|\{n\}\rangle_A\}$. However, zero-eigenvalue states of the boson parity operator $e^{i\pi \sum_k a_k^\dagger a_k}$ may exist:

$$e^{i\pi \sum_k a_k^\dagger a_k}|\phi^\pm\rangle = 0 \cdot |\phi^\pm\rangle \qquad (11)$$

Non-trivial solutions require vanishing determinant of the coefficient matrix in:

$$_A\langle\{m\}|e^{i\pi \sum_k a_k^\dagger a_k}|\phi^\pm\rangle = \sum_{\{n\}} D_{\{m\},\{n\}} c_{\{n\}}^\pm = 0 \qquad (12)$$

with matrix elements

$$D_{\{m\},\{n\}} = \prod_{k=1}^N D_{m_k, n_k}, \quad D_{m_k, n_k} = \sum_{j_k=0}^{\min[m_k, n_k]} \frac{e^{-2q_k^2} \sqrt{m_k! n_k!} (2q_k)^{m_k + n_k - 2j_k}}{(-1)^{j_k} (m_k - j_k)!(n_k - j_k)! j_k!}.$$

Truncating all modes to maximum occupation $M_{tr}$ (matrix dimension $M = M_{tr} + 1$), the coefficient matrix is a tensor product:

$$\left(D_{\{m\},\{n\}}\right)_{M \times M} = \bigotimes_{k=1}^N \left(D_{m_k, n_k}\right)_{M \times M} \qquad (13)$$

Its determinant is:

$$\left|\left(D_{\{m\},\{n\}}\right)_{M \times M}\right| = \prod_{k=1}^N |D^k|^M = \prod_{k=1}^N \left(e^{-2Mq_k^2}\right)^M = e^{-2M^2 \sum_{k=1}^N q_k^2} \qquad (14)$$

For sufficiently large M:

$$\left|\left(D_{\{m\},\{n\}}\right)_{M\times M}\right| = e^{-2M^2 \sum_k q_k^2} = 0 \qquad (15)$$

satisfying the non-trivial solution condition. At degeneracy:

$$E^+ = E^- = \left(\sum_k \omega_k m_k\right)_{\{m\}} - \sum_k \omega_k q_k^2 = E^{eo} \qquad (16)$$

$$|\phi^\pm\rangle = |\phi^{eo}\rangle \qquad (17)$$

This Δ-independent solution shows that all degeneracy points between parity energy spectra $E^+$ and $E^-$ lie on the degenerate energy line (16). Figure 1 illustrates for single-mode SBM: (a) Degeneracy points on $H_0$'s eigenvalues $E^{eo}$ (green); (b) Degeneracy count per quantum number n: no degeneracy at $E_0^\pm$, n degeneracies at $E_n^\pm$. The degeneracy count is topologically invariant for Δ ≠ 0, implying a non-degenerate ground state under Abel-group diagonalization.

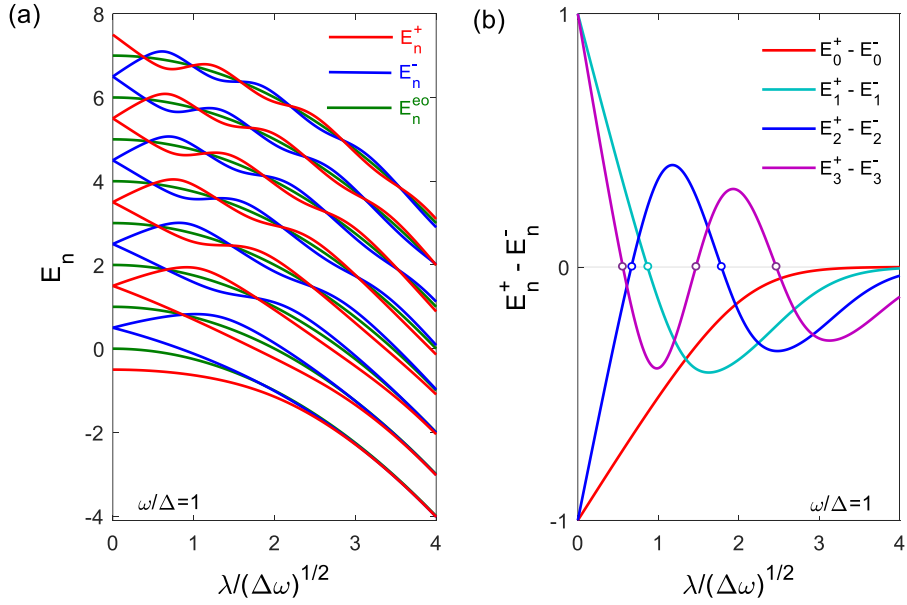

**Fig.1** single-mode SBM: (a) Degeneracy points on $H_0$'s eigenvalues $E^{eo}$ (green); (b) Degeneracy count per quantum number n: no degeneracy at $E_0^\pm$, n degeneracies at $E_n^\pm$.

## IV. Ground State Energy vs. Degenerate Energy in SBM

From (2), the ground state energy satisfies:

$$E_{gs} = min\{E_{min}^+, E_{min}^-\}. \tag{18}$$

For the auxiliary Hamiltonian $\tilde{H} = H^+ \oplus H^-$, $E_{min} = min\{E_{min}^+, E_{min}^-\}$.

Applying Rayleigh quotient to $2H_0 = (H^+ + H^-) \oplus (H^+ + H^-)$:

$$E_{min}^+ + E_{min}^- \geq \eta_{min}(2H_0) = 2E_{min}^{eo}. \tag{19}$$

Equality holds only when Δ = 0! For Δ ≠ 0, $E_{min}^+ \neq E_{min}^-$, yielding the strict inequality:

$$E_{gs} = min\{E_{min}^+, E_{min}^-\} < E_{min}^{eo} = -\sum_k \omega_k q_k^2. \tag{20}$$

Thus for Δ ≠ 0, $E_{gs} < E_{min}^{eo}$ and $\boldsymbol{E_{min}^+ \neq E_{min}^-}$, meaning tunneling lifts any Δ=0 degeneracy, giving the ground state definite parity. Consequently, magnetization vanishes:

$$\langle \sigma_z \rangle = \langle \psi | U^{-1} \sigma_z U | \psi \rangle = (0, \langle \phi^- |) \sigma_x \begin{pmatrix} 0 \\ |\phi^-\rangle \end{pmatrix} = 0. \tag{21}$$

Therefore, the magnetization spontaneous breaking signature of SBM QPT—transition from delocalized ($\langle \sigma_z \rangle$=0) to localized ($\langle \sigma_z \rangle$≠0) phase—cannot occur.

The degeneracy condition (16) applies universally. The lowest degenerate energy is:

$$E_{min}^{eo} = \begin{cases} -\lambda^2/4\omega & \text{(single mode)} \\ -\sum_{k=1}^N \lambda_k^2/4\omega_k & \text{(discrete multimode)} \\ -\dfrac{\alpha\omega_c}{2s} & \text{(continuum}, J(\omega) = 2\pi\alpha\omega_c^{1-s}\omega^s) \end{cases} \tag{22}$$

Figure 2 compares GPA-calculated ground state energy [17] with $E_{min}^{eo}$, confirming $E_{gs} < E_{min}^{eo}$.

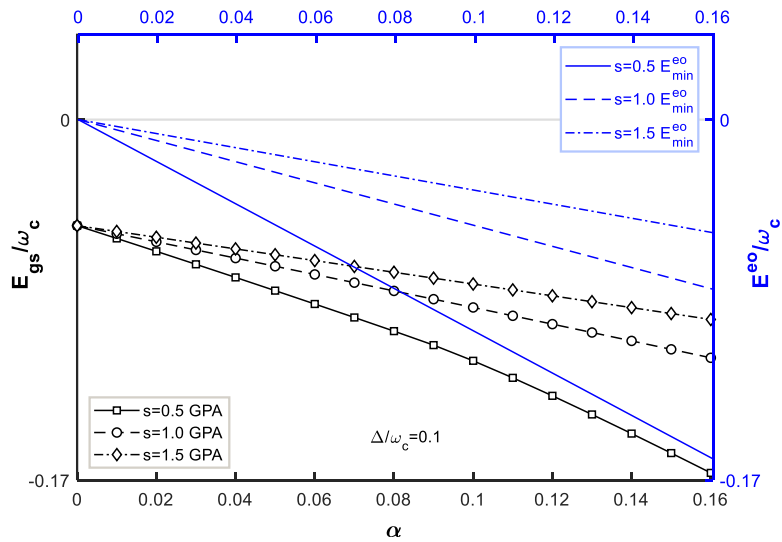

**Fig.2** compares GPA-calculated ground state energy [17] with E_min_eo.

Given the proven non-degenerate ground state with $\langle\sigma_z\rangle \equiv 0$, why have numerical methods consistently reported a "QPT"?

## V. Energy Level Degeneracy from Boson Parity Breaking and Critical Point

With computational accuracy $10^{-\Omega}$, the degeneracy condition is numerically satisfied when $\left|(D_{\{m\},\{n\}})_{M\times M}\right| < 10^{-\Omega}$. From (14):

$$2M^2 \sum_k q_k^2 \geq \ln(10^\Omega) \tag{23}$$

For continuous spectrum

$$J(\omega) = \pi \sum_k \lambda_k^2 \delta(\omega - \omega_k) = 2\pi\alpha\omega_c^{1-s}\omega^s \Rightarrow \sum_k \lambda_k^2 \delta(\omega - \omega_k) = 2\alpha\omega_c^{1-s}\omega^s,$$

we have:

$$\sum_k q_k^2 = \sum_k \frac{\lambda_k^2}{4\omega_k^2} = \int_\epsilon^{\omega_c} \frac{\sum_k \lambda_k^2 \delta(\omega - \omega_k)}{4\omega^2} d\omega = \frac{\omega_c^{1-s}}{2}\int_\epsilon^{\omega_c} \omega^{s-2} d\omega$$

For s < 1 (sub-Ohmic):

$$\sum_k q_k^2 = \frac{\omega_c^{1-s}}{2}\int_\epsilon^{\omega_c} \omega^{s-2} d\omega = \frac{\alpha\left(\left(\frac{\omega_c}{\epsilon}\right)^{1-s} - 1\right)}{2(1-s)}$$

Setting $\omega_c = 1$ and $\eta = \Omega/M^2$, Eq. (23) gives:

$$\alpha \geq \eta \ln(10)\frac{s-1}{1-\epsilon^{s-1}} \tag{24}$$

The parity-breaking critical point is defined as:

$$\alpha_c = \eta \ln(10)\frac{s-1}{1-\epsilon^{s-1}} \tag{25}$$

This is a finite-precision, finite-truncation critical point (not a true phase transition). For given accuracy $\Omega$ and truncation M, boson parity breaks numerically when $\alpha > \alpha_c$. Figure 3 shows excellent agreement between Eq. (25) and NRG/QMC "QPT" critical points: (a) For $\Delta/\omega_c = 0.1$, Eq. (25) (blue) matches NRG (black squares) and QMC (red circles) [PRL 102, 030601 (2009)]; (b) For $\Delta/\omega_c = 10^{-2}, 10^{-3}, 10^{-4}, 10^{-5}$, Eq. (25) (lines) matches NRG (symbols). Thus, the "critical point" α_c reflects computational limitations, not intrinsic ground state physics.

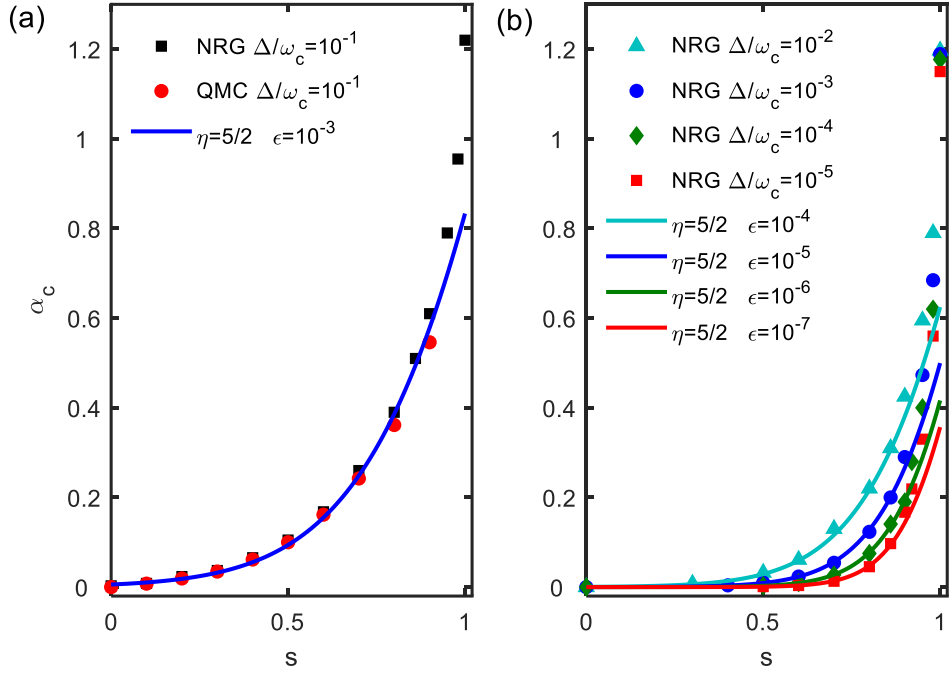

**Fig. 3** The curve of the bosonic parity-breaking critical point $\alpha_c \sim s$ reproduces the SBM phase diagram obtained by QMC and NRG methods. **(a)** Comparison of Eq. (25) with NRG and QMC results at $\Delta/\omega_c = 10^{-1}$: The blue solid line represents the result from Eq. (25), black squares denote NRG results, and red circles denote QMC results. NRG and QMC data are taken from Andre Winter, Heiko Rieger, Matthias Vojta, et al., PRL **102**, 030601-3 (2009). **(b)** Comparison of Eq. (25) with NRG results at $\Delta/\omega_c = 10^{-2}, 10^{-3}, 10^{-4}, 10^{-5}$: The four solid lines all represent results from Eq. (25), while triangles, circles, diamonds, and squares mark the numerical NRG results.

## VI. Conclusion

The agreement between NRG/QMC "phase diagrams" and Eq. (25) is not coincidental—it indicates that reported SBM phase transitions correspond to finite-accuracy parity breaking, not intrinsic ground state symmetry breaking. We have rigorously proven: (1) The SBM ground state is non-degenerate with $\langle \sigma_z \rangle \equiv 0$; (2) Numerical "QPTs" originate from computational truncation/accuracy effects captured by Eq. (25); (3) Genuine SBM quantum phase transitions arise from non-Abelian group diagonalization [18] producing non-equilibrium steady states, detailed in our forthcoming work.

### References

[1] A. J. Leggett et al., Rev. Mod. Phys. 59, 1(1987).
[2] R. Bulla, N.-H. Tong, M. Vojta, Phys. Rev. Lett. 91, 170601 (2003).

[3] K. L. Hur et al., Phys. Rev. Lett. 99, 126801 (2007).
[4] A. Winter et al., Phys. Rev. Lett. 102, 030601 (2009).
[5] R. Bulla et al., Phys. Rev. B 71, 045122 (2005).
[6] M. Vojta et al., Phys. Rev. Lett. 94, 070604 (2005); 102, 249904(E) (2009).
[7] F. B. Anders et al., Phys. Rev. Lett. 98, 210402 (2007).
[8] R. Bulla et al., Rev. Mod. Phys. 80, 395 (2008).
[9] M. Vojta et al., Phys. Rev. B 81, 075122 (2010).
[10] S. Florens et al., Phys. Rev. B 84, 155110 (2011).
[11] M. Cheng et al., Phys. Rev. B 80, 165113 (2009).
[12] A. Alvermann, H. Fehske, Phys. Rev. Lett. 102, 150601 (2009).
[13] H. Wong, Z.-D. Chen, Phys. Rev. B 77, 174305 (2008).
[14] C. Guo et al., arXiv:1110.6314 (2011).
[15] Z. Lü, H. Zheng, Phys. Rev. B 75, 054302 (2007).
[16] C. Zhao et al., Phys. Rev. E 84, 011114 (2011).
[17] A. W. Chin et al., Phys. Rev. Lett. 107, 160601 (2011).
[18] Tao Liu, "Irreversible Diagonalization of Mechanical Quantities and the EPR Paradox", arXiv:2409.15379 (2024).